\documentclass[prd,a4paper]{revtex4}

\usepackage{graphicx}
\usepackage{subfigure,amssymb}

\newcommand{\nco}{\newcommand}

\nco{\be}[1]{\begin{equation}\mbox{$\label{#1}$}}
\nco{\bea}[1]{\begin{eqnarray} \mbox{$\label{#1}$}}
\nco{\Section}[2]{\section{#2}\label{#1}}
\nco{\Bibitem}[1]{\bibitem{#1}}
\nco{\Label}[1]{\label{#1}}

\nco{\eea}{\end{eqnarray}}
\nco{\ee}{\end{equation}}

\nco{\bdm}{\begin{displaymath}}
\nco{\edm}{\end{displaymath}}
\nco{\dpsty}{\displaystyle}
\nco{\bc}{\begin{center}}
\nco{\ec}{\end{center}}
\nco{\ba}{\begin{array}}
\nco{\ea}{\end{array}}
\nco{\bab}{\begin{abstract}}
\nco{\eab}{\end{abstract}}
\nco{\btab}{\begin{tabular}}
\nco{\etab}{\end{tabular}}
\nco{\bit}{\begin{itemize}}
\nco{\eit}{\end{itemize}}
\nco{\ben}{\begin{enumerate}}
\nco{\een}{\end{enumerate}}
\nco{\bfig}{\begin{figure}}
\nco{\efig}{\end{figure}}

\nco{\arreq}{&\!=\!&}
\nco{\arrmi}{&\!-\!&}
\nco{\arrpl}{&\!+\!&}
\nco{\arrap}{&\!\!\!\approx\!\!\!&}
\nco{\non}{\nonumber}

\def\lsim{\; \raise0.3ex\hbox{$<$\kern-0.75em
      \raise-1.1ex\hbox{$\sim$}}\; }
\def\gsim{\; \raise0.3ex\hbox{$>$\kern-0.75em
      \raise-1.1ex\hbox{$\sim$}}\; }

\nco{\DOT}{\hspace{-0.08in}{\bf .}\hspace{0.1in}}
\nco{\Laada}{\hbox {$\sqcap$ \kern -1em $\sqcup$}}
\nco\loota{{\scriptstyle\sqcap\kern-0.55em\hbox{$\scriptstyle\sqcup$}}}
\nco\Loota{{\sqcap\kern-0.65em\hbox{$\sqcup$}}}
\nco\laada{\Loota}
\nco{\qed}{\hskip 3em \hbox{\BOX} \vskip 2ex}

\nco{\real}{{\rm I \! R}}
\nco{\Z}{{\sf Z \!\!\! Z}}
\nco{\complex}{{\rm C\!\!\! {\sf I}\,\,}}
\def\bigid{\leavevmode\hbox{\small1\kern-3.8pt\normalsize1}}
\def\id{\leavevmode\hbox{\small1\kern-3.3pt\normalsize1}}
\nco{\slask}{\!\!\!/}
\nco{\bis}{{\prime\prime}}
\nco{\pa}{\partial}
\nco{\na}{\nabla}
\nco{\ra}{\rangle}
\nco{\la}{\langle}
\nco{\goto}{\rightarrow}
\nco{\swap}{\leftrightarrow}

\nco{\EE}[1]{ \mbox{$\cdot10^{#1}$} }
\nco{\abs}[1]{\left|#1\right|}
\nco{\at}[2]{\left.#1\right|_{#2}}
\nco{\norm}[1]{\|#1\|}
\nco{\abscut}[2]{\Abs{#1}_{\scriptscriptstyle#2}}
\nco{\vek}[1]{{\rm\bf #1}}
\nco{\integral}[2]{\int\limits_{#1}^{#2}}
\nco{\inv}[1]{\frac{1}{#1}}
\nco{\dd}[2]{{{\partial #1}\over{\partial #2}}}
\nco{\ddd}[2]{{{{\partial}^2 #1}\over{\partial {#2}^2}}}
\nco{\dddd}[3]{{{{\partial}^2 #1}\over
    {\partial #2 \partial #3}}}
\nco{\dder}[2]{{{d #1}\over{d #2}}}
\nco{\ddder}[2]{{{d^2 #1}\over{d {#2}^2}}}
\nco{\dddder}[3]{{d^2 #1}\over
    {d #2 d #3}}
\nco{\dx}[1]{d\,^{#1}x}
\nco{\dy}[1]{d\,^{#1}y}
\nco{\dz}[1]{d\,^{#1}z}
\nco{\dl}[1]{\frac{d\,^{#1}l}{(2\pi)^{#1}}}
\nco{\dk}[1]{\frac{d\,^{#1}k}{(2\pi)^{#1}}}
\nco{\dq}[1]{\frac{d\,^{#1}q}{(2\pi)^{#1}}}

\nco{\bfT}{{\bf T }}
\def\GeV{{\rm GeV}}
\def\MeV{{\rm\ MeV}}

\nco{\cA}{{\cal A}}
\nco{\cB}{{\cal B}}
\nco{\cD}{{\cal D}}
\nco{\cE}{{\cal E}}
\nco{\cG}{{\cal G}}
\nco{\cH}{{\cal H}}
\nco{\cL}{{\cal L}}
\nco{\cO}{{\cal O}}
\nco{\cT}{{\cal T}}
\nco{\cN}{{\cal N}}
\nco{\cR}{{\cal R}}
\nco{\rvac}[1]{|{\cal O}#1\rangle}
\nco{\lvac}[1]{\langle{\cal O}#1|}
\nco{\rvacb}[1]{|{\cal O}_\beta #1\rangle}
\nco{\lvacb}[1]{\langle{\cal O}_\beta #1 |}
\nco{\bb}{\bar{\beta}}
\nco{\bt}{\tilde{\beta}}
\nco{\ctH}{\tilde{\cal H}}
\nco{\chH}{\hat{\cal H}}
\nco{\1}{\aa}
\nco{\2}{\"{a}}
\nco{\3}{\"{o}}
\nco{\4}{\AA}
\nco{\5}{\"{A}}
\nco{\6}{\"{O}}
\nco{\al}{\alpha}
\nco{\g}{\gamma}
\nco{\Del}{\Delta}
\nco{\e}{\textrm{e}}
\nco{\eps}{\epsilon}
\nco{\lam}{\lambda}
\nco{\Om}{\Omega}
\nco{\ve}{\varepsilon}
\nco{\mn}{{\mu\nu}}
\nco{\vp}{\varphi}

\nco{\rf}[1]{(\ref{#1})}
\nco{\nn}{\nonumber \\*}
\nco{\bfB}{\bf{B}}
\nco{\bfv}{\bf{v}}
\nco{\bfx}{\bf{x}}
\nco{\bfy}{\bf{y}}
\nco{\vx}{\vec{x}}
\nco{\vy}{\vec{y}}
\nco{\oB}{\overline{B}}
\nco{\oI}{\overline{I}}
\nco{\oR}{\overline{R}}
\nco{\rar}{\rightarrow}
\nco{\ti}{\times}
\nco{\slsh}{\hskip-5pt/}
\nco{\sm}{Standard~Model~}
\nco{\MP}{M_{\rm Pl}}
\nco{\mpl}{M_{\rm Pl}}
\nco{\tp}{t_{\rm Pl}}

\nco{\pmin}{p_{\rm min}}
\nco{\pmax}{p_{\rm max}}
\nco{\fo}{f_0}
\nco{\foi}{f_{0,i}\,}
\nco{\fop}{f_0^P}
\nco{\fou}{f_0^U}

\nco{\eff}{{\rm eff}}
\nco{\MT}{M_{\rm T}}
\nco{\ML}{M_{\rm L}}
\nco{\kk}{\vek{k}}
\nco{\pp}{{\rm p}}
\nco{\pt}{\partial_t}
\nco{\half}{{1\over 2}}
\nco{\w}{\omega}
\nco{\uhat}{\hat{U}_\w}

\nco{\etal}{\mbox{\it et al.}}
\nco{\ie}{{\it i.e. }}
\nco{\eg}{{\it e.g. }}
\nco{\trh}{T_{\rm RH}}
\nco{\ad}{{a'\over a}}
\nco{\bd}{{b'\over b}}
\nco{\Rd}{{R'\over R}}
\nco{\diag}{{\textrm{diag}}}
\nco{\mato}[1]{\tilde{#1}}
\nco{\sech}{\textrm{sech}}
\nco{\I}{\textrm{I}}
\nco{\II}{\textrm{II}}
\nco{\III}{\textrm{III}}
\nco{\vev}[1]{\langle #1 \rangle}
\nco{\hyp}{\,\; F_{1{\hskip -16pt}2}{\hskip 11pt}}
\nco{\brhom}{\overline{\rho}_M}
\nco{\brho}{\overline{\rho}}
\nco{\rhob}{\overline{\rho}}
\nco{\Pb}{\overline{P}}
\nco{\bH}{\overline{H}}
\nco{\ep}{{1+4\eps}}

\begin{document}

\title{Cosmological significance of one-loop effective gravity}

\author{Dom\`enec Espriu}
\affiliation{DECM and CER for Astrophysics, Particle Physics and Cosmology, Universitat de Barcelona,
Diagonal, 647, 08028 Barcelona, SPAIN}
\email[]{espriu@ecm.ub.es}
\author{Tuomas Multam\"aki}
\affiliation{NORDITA, Blegdamsjev 17, DK-2100, Copenhagen, DENMARK}
\email[]{tuomas@nordita.dk}
\author{Elias C. Vagenas}
\affiliation{Nuclear and Particle Physics Section, Physics Department, University of Athens,
GR-15771 Athens, GREECE}
\email[]{evagenas@phys.uoa.gr}

\date{}

\begin{abstract}
We study the one-loop effective action for gravity in a cosmological setup
to determine possible cosmological effects of quantum corrections to Einstein theory.
By considering the effect of
the universal non-local terms in a toy model, we show that they can play an important role
in the very early universe. We find that during inflation, the non-local terms
are significant, leading to deviations from the standard inflationary expansion.

\end{abstract}

\preprint{UB-ECM-PF-05/04, NORDITA-2005-20}

\maketitle

\section{Introduction}

Quantum corrections to Einstein action have received the attention of numerous authors for
quite a long time.
In the original works \cite{veltman}, the divergent structure
of the theory was determined.
It was seen that the theory required at the one-loop level ${\cal O}(\cR^2)$ counter terms
(albeit it turned out to be
finite on shell if matter was absent). These divergences will actually
be important for our discussion since,  by unitarity arguments, they determine the non-local
part of the one-loop effective action.

Although it was soon found out \cite{veltman} that higher order loops required more 
and more counter terms,
making Einstein gravity non-renormalizable, not even on-shell, this does not mean that
one-loop corrections
are useless. In fact in a regime of small curvatures
one-loop corrections will certainly dominate over two-loop corrections and so on. As a matter of fact
this is not
very different from the familiar expansion in chiral perturbation theory in powers of
$(p^2/16\pi f_\pi)^2$ (see e.g. \cite{gasser}).

Within this philosophy, the one-loop effective action for Einstein gravity has received
quite a lot of attention recently \cite{burgess}. For instance, the quantum corrections to Newton
law have been considered by a number of authors \cite{various}, and after some controversy
the correct result has been found. Quantum mechanics gives a correction
to the classical result of ${\cal O}(1/r^3)$ and positive; that is, at long distances
gravity is more attractive that what Newton law predicts. Of course the correction
vanishes anyhow at large values of $r$ and it is accompanied by a very small coefficient,
so for terrestrial or astronomical purposes this correction
is most likely irrelevant and unlikely to be tested ever.

However, the situation may be more promising from a cosmological view point. There is a
cumulative effect of gravity and, given a fixed density of energy, the integration of
this effect over large volumes could give an observable signal. This is, in short, the 
possibility that
 we would like to tentatively
study in this work.

We shall consider a flat Friedman-Robertson-Walker metric background and determine the
effect of quantum corrections on the cosmological evolution of the scale factor.
We shall concentrate here on the consequences of the non-local terms that necessarily appear
in the effective action due to unitarity considerations. It is a well known fact from
pion physics that these terms dominate al large distances, but somewhat surprisingly they
do not appear to have been considered before in the present context (with one exception to
which we shall turn below).
Somewhat to our surprise, the effect of these quantum corrections turns out to be relevant.

In the present, exploratory, paper we shall not consider the full one-loop effective action,
but shall limit ourselves to a toy model where only the scalar curvature is included
in the effective action and neglect the full Ricci or Riemann tensors. We shall also make for
simplicity a number of additional approximations that we shall discuss in more detail below.
The action of our toy model up to $\cR^2$ terms is
\be{action}
S = \int dx \sqrt{-g}\ \Big(\kappa^2\cR+\alpha\cR\ln(\nabla^2/\mu^2)\cR+\beta\cR^2\Big)+S_M
\ee
where $\kappa^2=1/(16\pi G)=M_{Pl}^2/(16\pi)$ and
$S_M$ includes the matter fields (and in particular the inflaton
sector of the theory). Note that the constants
$\alpha$ and $\beta$ are dimensionless. The expansion parameter is curvature
, $\cR$, which in the approximation $\dot{H}=0$ is in a simple relation
to the Hubble parameter, $\cR\sim H^2$. Hence, when $H/\mpl\ll 1$ \eg at
present times, local higher order terms can be neglected.
The values of $\alpha$ and $\beta$ are on a different theoretical footing.
$\alpha$ is entirely determined from the analytical structure of quantum 
corrections induced by the lowest
dimensional term; in other words,
its value is uniquely determined once one insists in the long distance description of
gravity being provided by Einstein theory. $\beta$, on the other hand, is model dependent.
Its value is fixed as a boundary condition upon integration of other degrees of freedom
that do not have been included in (\ref{action}). Furthermore, $\beta$ is  scale dependent
so as to cancel the (local) $\ln \mu^2$ dependence appearing normalizing the log
in (\ref{action}).

We neglect higher orders in the expansion in derivatives of the metric. A power counting
can be established here \cite{counting}, in parallel to what is done 
in chiral perturbation theory. We
shall therefore work with a precision where only up to four derivatives of the metric
need to be included.

\section{Determination of the equations of motion}

We split the action (ignoring the matter part for now) into three parts and
redefine the constants for
convenience
\bea{act2}
S & = & \kappa^2 \Big(\int dx\sqrt{-g}\ \cR+\tilde{\alpha}
\int dx\sqrt{-g}\ \cR\ln(\nabla^2/\mu^2)\cR+\tilde{\beta}\int dx\sqrt{-g}\
\cR^2\Big)\nonumber\\
& \equiv & \kappa^2\Big(S_1+\tilde{\alpha}S_2+\tilde{\beta}S_3\Big),
\eea
where $\tilde{\alpha}=\kappa^{-2}\alpha,\ \tilde{\beta}=\kappa^{-2}\beta$. The dimensionful
constant $\mu$ is actually a subtraction scale that is required for dimensional consistency. The
coupling $\tilde{\beta}$ is $\mu$ dependent in such a way that the total action $S$ 
is $\mu$-independent.
In conformal time, $dt=a\ d\tau$, we have
\be{conft}
g_{\mu\nu}=a^2(\tau)\eta_{\mu\nu},\ \cR=-6 {a''(\tau)\over a^3(\tau)},\
\sqrt{-g}=a^4(\tau).
\ee
Variation of the local action is straightforward and gives the following results
\bea{s1var}
{\delta S_1\over\delta a(\tau)} & = & -12 a''(\tau)\\
{\delta S_3\over\delta a(\tau)} & = & 72\Big(-3{(a'')^2\over a^3}-
4 {a'a'''\over a^3}+6{(a')^2a''\over a^4}+
{a^{(4)}\over a^2}\Big).
\eea

The variation of $S_2$ is more delicate and requires
some discussion. The d'Alembertian in conformal space is related to the Minkowski space operator by
\cite{birrelldavies}
\be{curvedboxed}
\nabla^2={a}^{-3}\Box\, a +\frac16 \cR.
\ee
To the precision we are working in the curvature expansion we can neglect the $\cR$-term
in the expansion of the d'Alembertian and commute the scale factor $a$ with the flat d'Alembertian;
therefore we set
\be{curvedboxed2}
\nabla^2=\Big({a\over a_0}\Big)^{-2}\Box,
\ee
Where $a_0=a(0)$. The rescaling (absorbable in $\tilde{\beta}$) ensures that
at $\tau=0$ the d'Alembertian matches with the Minkowskian one.
In fact, from now on we will set $a_0=1$ for simplicity.

We can now separate $S_2$ into a local and a non-local piece
\bea{s22}
S_2 & = & \int dx\sqrt{-g}\ \Big(-2\cR\ln(a)\cR+\cR\ln(\Box/\mu^2)\cR\Big)\nonumber\\
& = & S_2^I+S_2^{II}.
\eea
Variating the first term $S_2^I$ is again straightforward
\bea{s2Ivar}
{\delta S_2^I\over\delta a(\tau)} & = & -72\Big\{2\Big(6\ln(a)-5\Big){(a')^2a''\over a^4}
+4\Big(1-2\ln(a)\Big){a'a'''\over a^3}+3\Big(1-2\ln(a)\Big){(a'')^2\over a^3}
+2\ln(a){a^{(4)}\over a^2}\Big)\nonumber\}.
\eea
We shall discuss the variation of $S_2^{II}$ next.

\subsection{Evaluation of the non-local contribution}
The genuinely non-local piece in $S_2$ is
\bea{s2II}
S_2^{II} & = & \int dx\sqrt{-g}\ \cR\ln(\Box/\mu^2)\cR\nonumber\\
& = & \int dx\sqrt{-g(x)}\cR(x)\int dy\sqrt{-g(y)}\ \langle x|\ln(\Box/\mu^2)|y\rangle\cR(y).
\eea
To evaluate this logarithmic term, we make use of the identity
$\ln(x)\approx -1/\epsilon+x^\epsilon/\epsilon$, valid when $\epsilon$ is small.
Using this, we see that we need to compute
$\epsilon^{-1}\langle x|(\Box/\mu^2)^\epsilon|y\rangle$,
that has the integral representation
\bea{intrep}
\epsilon^{-1}\langle x|(\Box/\mu^2)^\epsilon|y\rangle
& = & {1\over\epsilon} 2\pi^2\mu^{-2\epsilon}
\int dk\ k^{2+2\epsilon}{1\over |x-y|}J_1(k|x-y|)\nonumber\\
& \sim & -8\pi^2\mu^{-2\epsilon}{1\over |x-y|^{4+2\epsilon}}
\eea
where $J_1$ is the Bessel function.
As we are only interested in the time evolution of the scale factor
we can integrate out the spatial dependence in Eq. (\ref{s2II})
absorbing the constants into
$\tilde{\alpha}$ leaving
\bea{intrep2}
S_2^{II} & = & \int d\tau\sqrt{-g(\tau)}\cR(\tau)\int d\tau'
\sqrt{-g(\tau')}\ \mu^{-2\epsilon}{1\over |\tau-\tau'|^{1+2\epsilon}}\cR(\tau')
\eea
The above integral is of course divergent; the divergence is local and is exactly cancelled
by the $-1/\epsilon$ in $\ln(x)\approx -1/\epsilon+x^\epsilon/\epsilon$. 
By taking this term into account
and taking $\epsilon\to 0$, (\ref{s2II}) can be easily numerically computed for a given background.

A final word of caution has to do with the causal conditions to be imposed on our Green
functions. We have been so far rather cavalier about that. When deriving evolution equations
(as opposed to $S$-matrix elements) one has to be careful with the causal conditions. The usual
Feynman rules lead to the so-called in-out effective action, relevant for $S$-matrix elements.
When interested in causal evolution one has to consider the in-in effective action \cite{jordan}.
This affects the causal definition of the Green function
$\Delta(x,y)\equiv\langle x|(\Box/\mu^2)^\epsilon|y\rangle $. If $\Delta_F,\Delta_D,
\Delta_+$ and $\Delta_-$ are the Feynman, Dyson, advanced and retarded Green functions, respectively,
the combination that actually appears in the in-in effective action
takes the matrix form (see e.g. \cite{in-in} for a clear discussion on this formalism)
\bea{inin}
(\cR_+\, \cR_-)\left(\matrix{\Delta_F & -\Delta_+\cr
                              \Delta_- & -\Delta_D}\right)
\left(\matrix{\cR_+ \cr
      \cR_- }\right),
\eea
where the subscripts denote positive and negative frequency, respectively.
This ensures that when taking a functional derivative w.r.t. $a(t)$, only earlier times
contribute to the evolution. This can be enforced by setting the limits in the time
integrals appropriately.

\subsection{Variation of the non-local contribution}
Now we can proceed with the variation of the non-local part
(for simplicity, we set  $\epsilon=0$ here; although the expressions are all ill-defined, 
we can replace
$\epsilon$ back easily)
\bea{s2IIvar}
{\delta S_2^{II}\over\delta a(\tau)} & = & 72 \int d\tau'\Big\{ {\delta\over\delta a(\tau)}
\Big(a(\tau')a''(\tau')\Big)
\int_{0}^{\tau'} d\tau'' {1\over \tau'-\tau''} a(\tau'')a''(\tau'')\Big\}\nonumber\\
& = & 72\Big\{2 a''\int_{0}^{\tau} d\tau' {1\over \tau-\tau'} a(\tau')a''(\tau')
+2 a' \partial_\tau\Big(\int_{0}^{\tau} d\tau' {1\over \tau-\tau'} a(\tau')a''(\tau')\Big)
+a \partial_\tau^2\Big(\int_{0}^{\tau} d\tau' {1\over \tau-\tau'} a(\tau')a''(\tau')\Big)\nonumber
\eea
The derivatives can be easily performed by integration by parts
\bea{dt}
\partial_\tau\Big(\int_{0}^{t} d\tau' {1\over \tau-\tau'} a(\tau')a''(\tau')\Big)
& = & {a_0 a''_0\over\tau}+\int_{0}^{\tau} d\tau' {1\over \tau-\tau'}\Big(a'(\tau')a''(\tau')
+a(\tau')a'''(\tau')\Big)\\
\partial^2_\tau\Big(\int_{0}^{\tau} d\tau' {1\over \tau-\tau'} a(\tau')a''(\tau')\Big) & = &
-{a_0 a''_0\over\tau^2}+{a_0'a_0''+a_0a_0'''\over\tau}\nonumber\\
& & + \int_{0}^{\tau} d\tau' {1\over \tau-\tau'}\Big(a''(\tau')^2+2 a'(\tau')a'''(\tau')
+a(\tau')a^{(4)}(\tau')\Big).
\eea
The poles at $\tau=0$ terms are an artifact arising from the fact that at $\tau=0$
we patch together Minkowski space and de-Sitter space by starting inflation at that
point. If this is done smoothly enough, the derivatives of the scale factor vanish
at that point (this is evidenced by the fact that all these terms 
contain derivatives of $a$ at $\tau=0$).
Hence, we disregard these terms in the following. We have checked that
modifying the matching has unobservable consequences in our results.

\section{Effects on inflation}
With the variated action, we can now look for solutions of
the resulting equation. In general, the equation is a complicated
integro-differential equation that is most suited for numerical
studies. However, we expect that the new terms may have an effect
during inflation when $\cR$ is large and $\dot{H}$ is small.

In conformal time, the scale factor grows during inflation as
\be{confinf}
a_I(\tau)={1\over 1-H \tau},
\ee
where $H$ is the inflationary Hubble rate as in $a=a_0\exp(Ht)$.
It is evident that in conformal time we must restrict the time interval
considered to $\tau<1/H$. Note that in order for this to be the solution
of the varied gravitational action (without the higher curvature terms)
one has to put in the appropriate matter terms into the action.
Assuming a simple generic inflationary model, one can relate the
the Hubble rate to the inflaton potential by $H^2=8\pi GV(\phi)/3$,
where $\phi$ is the value of the inflaton during inflation
(approximated with a constant).
Here we consider $H$ to be a constant and study how the new terms
affect the inflationary expansion.
In a more realistic treatment left for future work, one
has to let the inflaton roll and solve the coupled system of
equations of the inflaton equation along with the modified
Einstein's equation.

We proceed by solving the varied gravitational action by a
perturbative approximation, \ie we consider the non-standard terms
as a correction to the standard inflationary solution. This procedure
is only valid as long as the correction is small compared to the unperturbed
solutions, which sets the limits for the validity of our approach.

For calculations it is useful to note the relation
\be{ansrel}
a_I'(\tau)=H a_I^2(\tau),
\ee
which implies that
\be{ansrel2}
a_I^{(n)}=n! H^na_I^{n+1}(\tau).
\ee

The $0$-th order equation, corresponding
to including the inflaton sector in the action, is
\be{zeroord}
-12a''+24 H^2a^3=0,
\ee
which Eq. (\ref{confinf}) is a solution of.

\subsection{Perturbing the $0$-th order solution}

By using the $0$-th order solution (\ref{confinf}) we see that the variation of $S_3$ vanishes
(this is not a general result, \ie this only happens with the $\cR^2$ term, other terms
of the form $\cR^n,\ n>2$ are non-zero). The variation of $S_2$ does not vanish and
after a straightforward substitution we get for each part
\bea{s2varans}
{\delta S_2^I\over\delta a(\tau)} & = & -1152 {H^4\over (1-H  \tau)^3}=-1152 H^4a_I^3\\
{\delta S_2^{II}\over\delta a(\tau)} & = & 72\Big\{4H^2a_I^3\Big(2\int_0^\tau\ 
d\tau'{1\over \tau-\tau'}H^2a_I^4(\tau')\Big)+
2Ha_I^2\Big(8\int_0^\tau\ d\tau'{1\over \tau-\tau'}H^3a_I^5(\tau')\Big)\nonumber\\
& & +a_I\Big(40\int_0^\tau\ d\tau'{1\over \tau-\tau'}H^4a_I^6(\tau')\Big)\Big\}.
\eea
Hence, the $1$st order equation of motion for $a(\tau)$ is
\bea{finalequ}
a''-2H^2a^3 & = & 12 \tilde{\alpha}\Big\{
-8 H^4a_I^3\nonumber\\
& &
+4H^4a_I^3\int_0^\tau\ d\tau'\,{a_I^4(\tau')\over \tau-\tau'}
+8H^4a_I^2\int_0^\tau\ d\tau'\,{a_I^5(\tau')\over \tau-\tau'}
+20H^4a_I\int_0^\tau\ d\tau'\,{a_I^6(\tau')\over \tau-\tau'}
\Big\}.
\eea
The r.h.s. of this equation (that depends on the unperturbed $0$-th order solution only) can
be regarded as a driving external force for the evolution equation of the conformal factor.

The integral terms can be explicitly computed (we now restore $\mu$ and $\epsilon$)
\be{intsimpl}
\mu^{-2\epsilon}\int_0^\tau\ d\tau'\,{a_I^n(\tau')\over (\tau-\tau')^{1+2\epsilon}}
=-{1\over\epsilon}(\tau\mu)^{-2\epsilon}F(n,1,1-\epsilon, H\tau),
\ee
where $F$ is the hypergeometric function. In the limit $\epsilon\rar 0$,
Eq. (\ref{intsimpl}) tends to
\bea{intsimpl2}
\mu^{-2\epsilon}\int_0^\tau\ d\tau'\,{a_I^n(\tau')\over (\tau-\tau')^{1+2\epsilon}}
& = & -{1\over\epsilon}F(n,1,1, H\tau)+\ln(\tau\mu)F(n,1,1, H\tau)+
\partial_\epsilon F(n,1,1-\epsilon, H\tau)\Big|_{\epsilon\rar 0}\nonumber\\
& = & -{1\over\epsilon}(1-H\tau)^{-n}+(1-H\tau)^{-n}\ln(\tau\mu)+
\partial_\epsilon F(n,1,1-\epsilon, H\tau)\Big|_{\epsilon\rar 0},
\eea
indicating the need to regulate the integrals by the addition of an
appropriate counter-term (this is the role of the $-1/\epsilon$ term appearing in the 
logarithm representation).
We are now ready to solve Eq. (\ref{finalequ}) numerically,
using the $0$-th order solution (\ref{confinf}) as an initial condition.

\subsection{Numerical analysis}

As we can see from the form of the inflationary expansion in conformal time, the
scale factor is singular at $\tau=1/H$. Furthermore,
if we wish that \eg $a_I/a_0\sim \e^{60}\sim 10^{26}$,
we must require that $1-H\tau\sim 10^{-26}$, \ie that $\tau$ is very close to the
singular point. It is hence useful for numerical work to do a change of variables,
$\e^s=1/(1-H\tau)$. In these coordinates the inflationary expansion is simply $a_I(s)=\e^s$
and the coordinate $s\in[0,\infty]$.
The equation of motion with the new dimensionless time coordinate is
(after regularization)
\be{finalequ3}
\e^{2s}a''+\e^{2s}a'-2a^3=12 \tilde{\alpha}H^2\Big(
-8 \e^{3s}+4\e^{3s}G(4,s)+8\e^{2s}G(5,s)+20\e^{s}G(6,s)\Big)
\ee
where we have defined $G(n,s)\equiv
\e^{ns}\ln({\mu\over H}(1-\e^{-s}))+
\partial_\epsilon F(n,1,1-\epsilon, 1-\e^{-s})\Big|_{\epsilon\rar 0}$ and
divided both sides by $H^2$.
In this form of the equation of motion, we note a number of interesting properties.
First of all we see that the coupling constant appears in a combination
$\tilde{\alpha}H^2$, indicating that the corrections to the standard
evolution become extremely small
at late times (recall that presently $H\sim 10^{-42}\GeV$ and that $\tilde\alpha$ 
is dimensionful and proportional to the inverse Planck mass). Secondly,
the arbitrariness in the choice of the scale is exhibited by the presence of
the $\mu/H$ term inside the logarithm. 
Finally, we note that the right hand side is singular
at $s=0\ (\tau=0)$ but this is only a logarithmic divergence and can be avoided
in numerical work by starting the calculation at some small non-zero value of $s$.

For numerical work, we need to estimate the scale of inflation, $V(\phi_{inf})\equiv V_0$,
and hence the Hubble parameter during inflation, $H^2=8\pi G V_0/3$.
As an absolute lower limit, the energy density in the inflaton at the end
of inflation must be enough to reheat the universe to a high enough
temperature for nucleosynthesis to occur ($T\sim 1\MeV$), but typical values 
can be much bigger than this,
up to the CMB normalization limit $V_0^{1/4}\lsim 10^{16}\GeV$, corresponding to
$H\sim 10^{13}\GeV$.

It should be stressed here that the model proposed in (\ref{action}) is not a realistic one
inasmuch as we only include $\cR^2$ terms and neglect other possible curvature contributions.
Our purpose here is to test in a simple setting whether these quantum contributions 
could be important at all.
Therefore, albeit the actual values of $\alpha$ are known (see e.g. \cite{dobado}),
after integration of gravitons and the rest of massless particles in the Standard Model 
(massive particles are irrelevant for this discussion as they do not provide logs), 
we provide numerical results for several values of $\alpha$ around the value 
$\vert \alpha\vert\sim 10^{-4}$, which is the natural order of
magnitude expected. A more accurate treatment would require the introduction of all
the Riemann curvature tensor components. It would also require to establish which 
particles are exactly massless (or have a mass much smaller than the inverse horizon radius
for that matter). 

There is also a built-in $\mu$ dependence in our results. As we have stressed, 
the results should be, in principle, $\mu$-independent (this is, incidentally, 
a point that is often overlooked in this type of analysis). The dependence on 
$\mu$ in the non-local piece is exactly compensated by
the (logarithmic) $\mu$ dependence of $\beta$. Of course we do not know the value of 
$\beta$ as it contains contributions from all modes that have been integrated out and 
that are not explicitly included in the lagrangian. We can, however, estimate the `natural'
scale for $\mu$; i.e. the one that minimizes higher order corrections. 

Taking into account that massive particles are integrated out and that they do not generate
logs, within a renormalization group approach it is natural to consider that the natural
scale is that of the lightest particle that has been integrated out. This is similar to what
is done in effective lagrangians for the strong interactions such as chiral lagrangians, 
where the optimal scale is somehow related to the scale of chiral symmetry breaking that
separates `light' degrees of freedom from `heavy' ones. In the Standard model we assume
this lightest mass to be 1 meV, corresponding to the neutrino. 

\begin{figure}
\subfigure[$\mu=M_{Pl}$.]{\includegraphics[width=8.5cm]{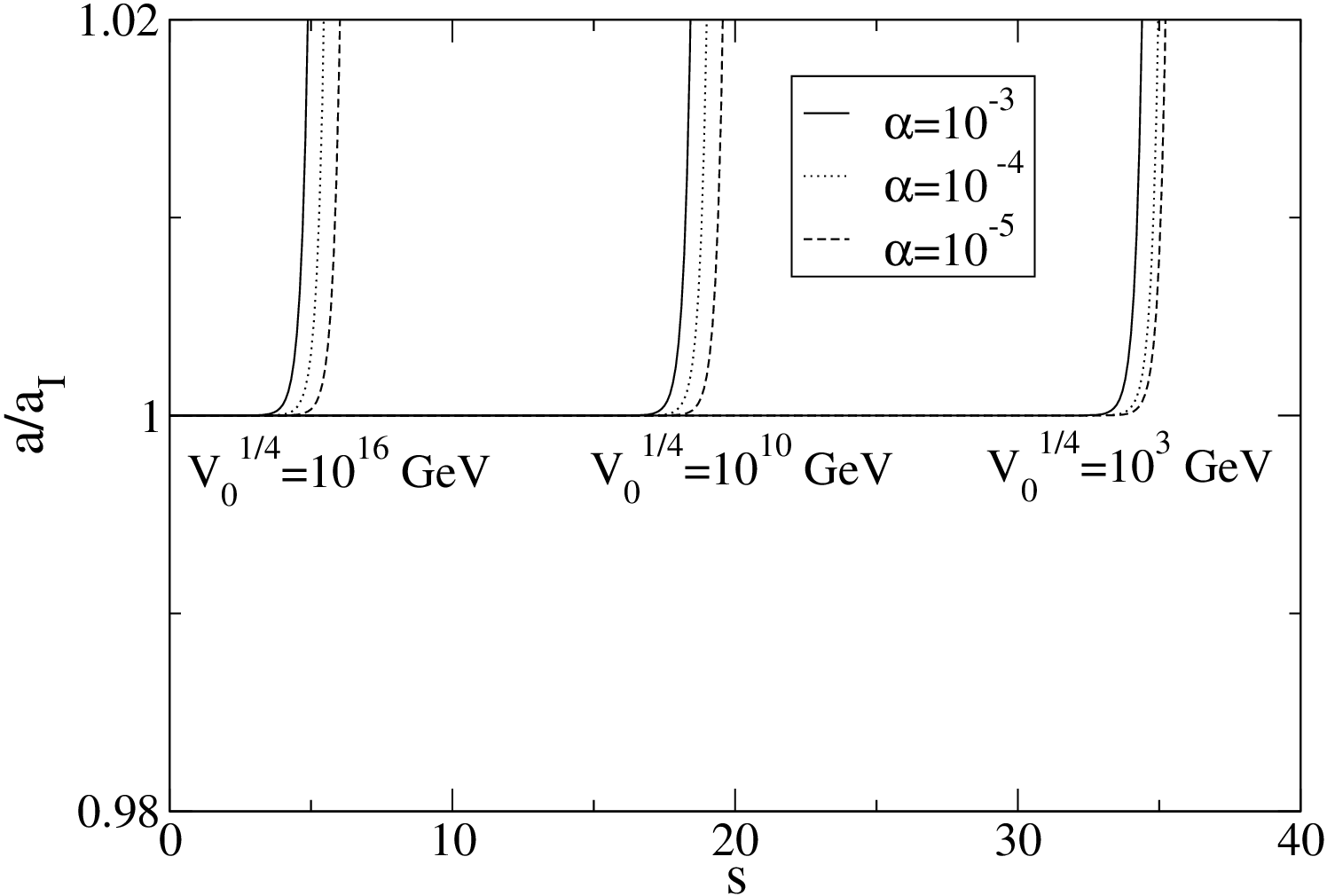}}
\subfigure[$\mu=1\,\textrm{meV}$.]{\includegraphics[width=8.5cm]{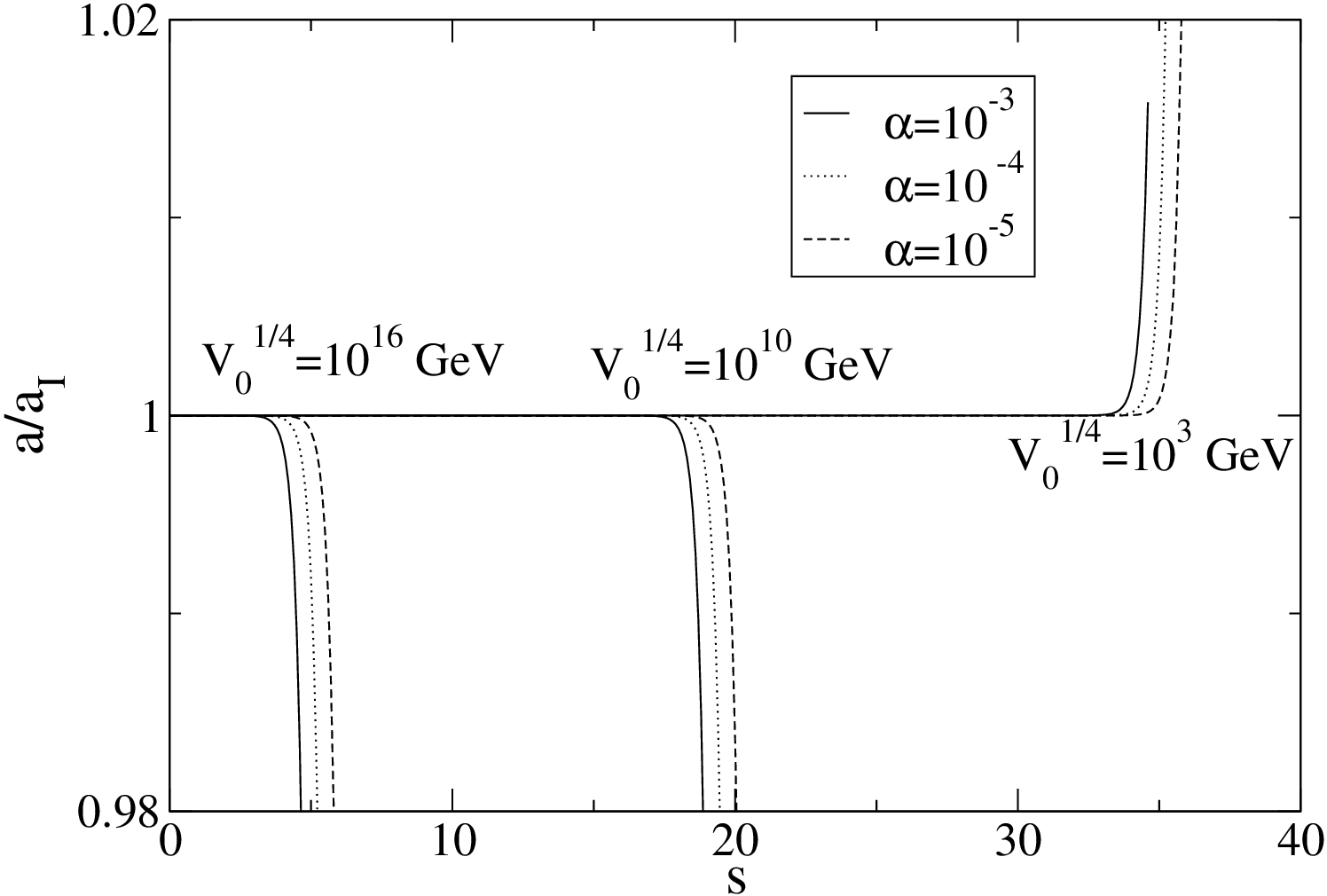}}
\caption{The scale factor relative to the inflationary expansion 
for different values of $\alpha,\ \mu$ and $V_0$.}\label{fig1}
\end{figure}

To understand the effect of all of the previous choices, we have solved
Eq. (\ref{finalequ3}) numerically for different values of $\alpha,\ V_0$ and $\mu$.
They are shown in Figs \ref{fig1}(a) and \ref{fig1}(b) as a ratio of the scale factor to the
inflationary expansion, $a/a_I$.

The normalization scales are chosen to represent the minimum and maximum values.
Note that here we only show results for positive $\alpha$. For negative alpha,
the curves simply turn in the opposite direction at the same value of $s$ and therefore
we choose not to shown them here.

From the figures one can see how a higher inflationary scale leads to deviations from
the inflationary expansion earlier than a lower scale. This is as expected since
the source term in Eq. (\ref{finalequ3}) is proportional to $H$ and hence to $V_0$.
Similarly, a larger $\alpha$ has the same effect. 
The effect of changing the normalization scale $\mu$ has a mixed effect. If $\mu=M_{Pl}$,
the $\ln(\mu/H)$-term in $G(s,n)$ is positive for all considered values of $V_0$. 
However, if $\mu=1\,\textrm{meV}$, the logarithmic term changes sign from negative to 
positive at $V_0\sim 2\times 10^3\,\GeV$ so that at large $V_0$, the source 
term is negative.
In the source term, the $-8e^{3s}$ term is subdominant compared to the other terms. The
hypergeometric function gives a negative contribution for all $n$ values considered here.
Hence, the the sign of the logarithmic term is crucial, as is clear from the figures.

\begin{figure}
\includegraphics[width=7cm]{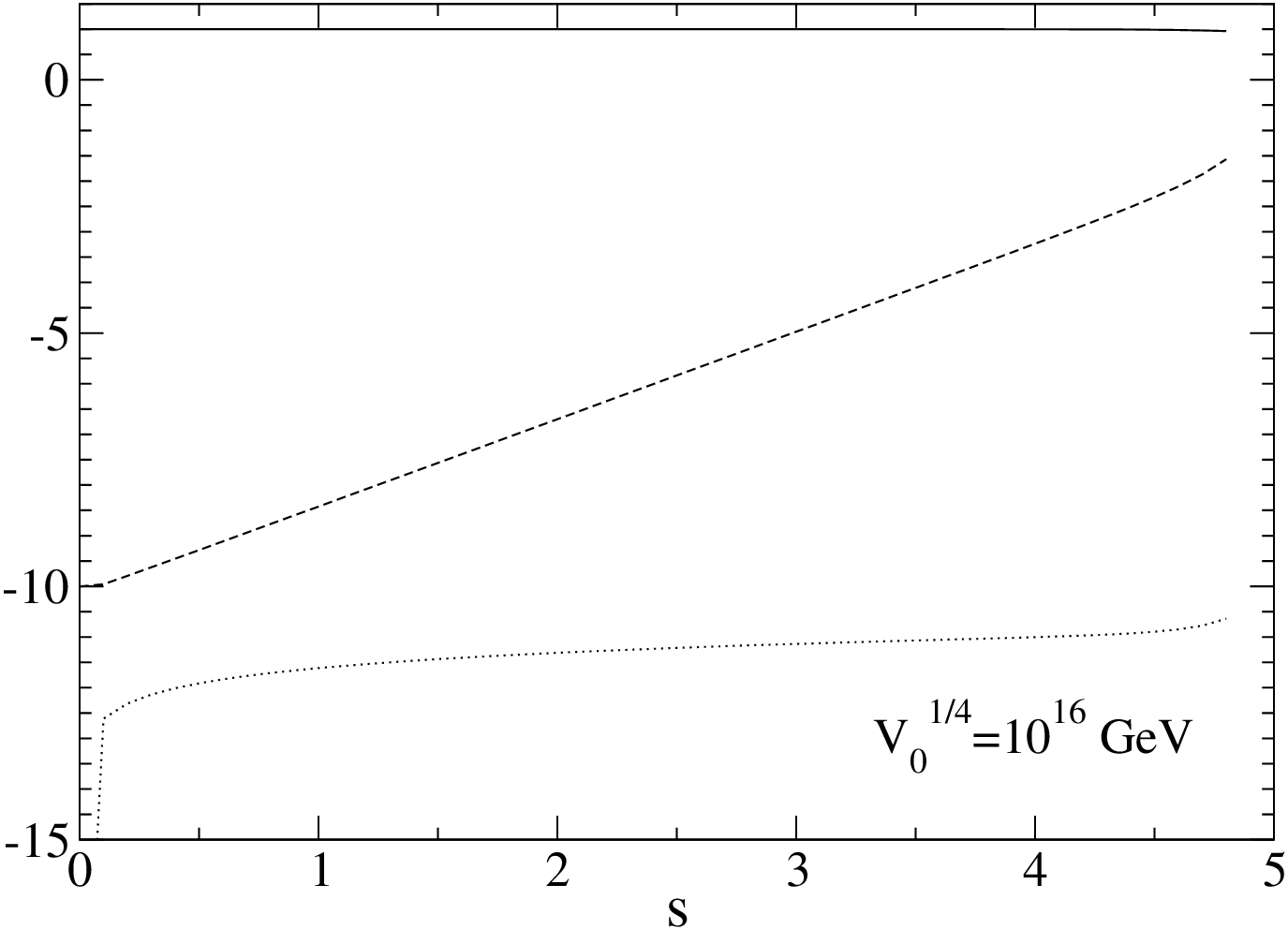}
\caption{The scale factor relative to $a_I$ (solid line), 
$\log_{10}(|2\mathcal{R_I}\ln(a)\mathcal{R_I}/\mathcal{R}|)$ (dotted line), 
$\log_{10}(|\mathcal{R_I}\ln(\Box/\mu^2)\mathcal{R_I}/\mathcal{R}|)$ (dashed line)
for $\alpha=10^{-3},\ \mu=1\,$meV and $V_0^{1/4}=10^{16}\GeV$.}\label{fig2}
\end{figure}

The relative contributions to the action arising from the logarithmic term is shown for
a particular case in
Fig. \ref{fig2}, along with the corresponding numerical solution. From the figure one 
can see that the perturbative approximation we are making is appropriate as the action
is still dominated by the Einstein term, $\mathcal{R}$. Note that at then end of
the corresponding calculation the relative scale factor has decreased to about
$0.96$. The new terms are also subdominant for the other choice of parameter values 
shown in Fig. \ref{fig1}.

\section{Conclusions}
In this paper we have considered the effects of quantum corrections to gravity in a somewhat 
simplified setting.
By considering the non-local terms that necessarily appear due to unitarity 
considerations in any effective lagrangian involving massless particles, that 
typically dominate long-distance physics, 
we have found that
during inflation the non-local effects are important and lead to deviations from 
the standard inflationary expansion. The effect is sizeable, as for typical inflationary 
parameter values the expansion rate is changed after only $10-20$ e-foldings.
The sign of the effect depends on the parameter values, and in particular on the sign
of $\alpha$ and size of the normalization scale $\mu$. Recalling that a change in $\mu$ 
is tantamount to a change in the coefficient of the local $\cR^2$ term, pinning down
the physical value for $\beta$ (and its proper $\mu$ dependence) is very important.

Taking into account that quantum corrections actually strengthen gravity at long distances,
we believe that in the physically relevant situation, inflation would be slowed down or halted
by the quantum corrections.

This type of effects have not, to our knowledge been considered before, except for the studies 
presented in \cite{tsamis}. Although no detailed numerics are presented in this reference, 
the authors
conclude that quantum effects slow inflation. Unfortunately the two approaches could hardly be
more different and hence comparison is hopelessly difficult. It would be nice to make 
a clear contact between the two approaches.

In doing the calculations, we have done a number approximations and simplifications
in order to see whether the quantum effects can be important. As this has proven to be so
in the toy model considered here, the effects of the non-local terms need to be studied
more carefully in a more realistic model. We certainly believe that this issue deserves 
further studies.

\section*{Acknowledgements}
The financial support of EUROGRID and ENRAGE European Networks is gratefully acknowledged.
D.E. would like to thank J. Garriga, A. Maroto, A. Roura and E. Verdaguer for discussions.
The work of E.C.V. is financially supported by the PYTHAGORAS II 
Project ``Symmetries in Quantum and Classical Gravity''
of the Hellenic Ministry of National Education and Religions.
E.C.V. would like to thank J. Russo and S. Odintsov for useful correspondences.
The work of D.E. is supported by project FPA2004-04582.



\end{document}